\documentclass[twoside,a4paper]{amsart}

\PassOptionsToPackage{pdfauthor={Vladimir V. Kisil},%
    pdftitle={Computation and Dynamics: Classical and Quantum},%
    pdfsubject={mathematics},%
    backref=page,%
  pdfkeywords={symmetries}
}{hyperref}
\IfFileExists{eulervm.sty}{\usepackage{eulervm}
}{}

\newtheorem{thm}{Theorem}[section]
\newtheorem{prop}[thm]{Proposition}
    
\theoremstyle{definition}
\newtheorem{example}[thm]{Example}
\newtheorem{conj}[thm]{Conjecture}
\theoremstyle{remark}
\newtheorem{rem}[thm]{Remark}
\RequirePackage[breaklinks=true,%
colorlinks=true,%
bookmarks=true,%
backref=page,%
pagebackref=true]{hyperref}
\usepackage[backrefs]{amsrefs}

\makeatletter
\PII{\relax}
\copyrightinfo{}{\relax}
\makeatother
\usepackage{graphicx}

\providecommand{\mod}{\mathrm{mod}\,}

\usepackage{amsfonts}
\providecommand{\comment}[1]{}
\providecommand{\Space}[3][]{\ensuremath{\mathbb{#2}^{#3}_{#1}{}}}

\providecommand{\wiki}[2]{\href{http://en.wikipedia.org/wiki/#1}{#2}}

\newcommand{\person}[1]{#1}
\begin{document}
\title[Computation and Dynamics: Classical and Quantum]{Computation and Dynamics:\\
 Classical and Quantum}
\author[Vladimir V. Kisil]%
{\href{http://www.maths.leeds.ac.uk/~kisilv/}{Vladimir V. Kisil}}
\thanks{On  leave from Odessa University.}
\thanks{The author acknowledges the support of the EPSRC Network on Semantics of 
Quantum Computation (EP/E006833/2).}

\address{%
School of Mathematics\\
University of Leeds\\
Leeds LS2\,9JT\\
UK
}

\email{\href{mailto:kisilv@maths.leeds.ac.uk}{kisilv@maths.leeds.ac.uk}}

\urladdr{\href{http://www.maths.leeds.ac.uk/~kisilv/}%
{http://www.maths.leeds.ac.uk/\~{}kisilv/}}

\begin{abstract}
  We discuss classical and quantum computations in terms of
  corresponding Hamiltonian dynamics.
\end{abstract}
\maketitle

\section{Introduction}
\label{sec:introduction}
It is well known that classical computations are modelled by abstract
\wiki{Turing_machine}{``machines''} first introduced in works of
\wiki{Emil_Post}{E.~Post} and \wiki{Alan_Turing}{A.~Turing}, see
\cite{KnuthACP250}*{\S~1.4.5 and \S~2.6} for historical notes and
further references.

We are going to demonstrate and exploit an explicit analogy between
the process of computation on such abstract machines and a Hamiltonian
dynamics of a particle in the phase space. We will use for this
purpose the Post machine since its description is simpler. In the
following we will call it simply the \emph{machine}.

\begin{figure}[htbp]
  \centering
  (a)\includegraphics[scale=.9]{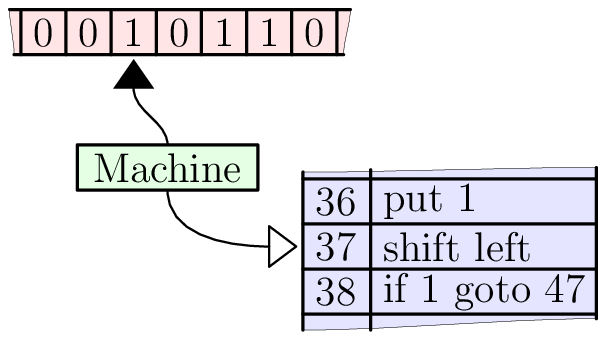}\hfill
  (b)\includegraphics[scale=.9]{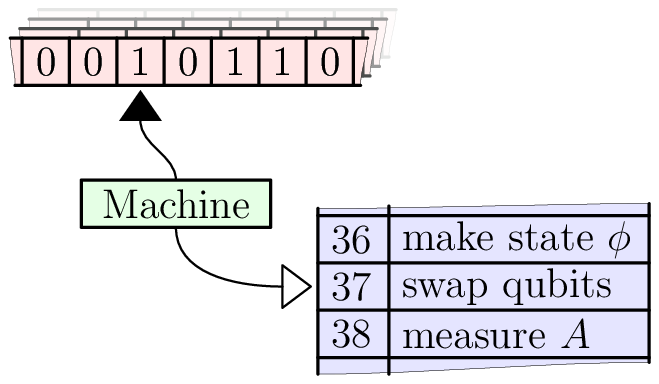}
  \caption[A symbolic representation of the Post machine]{(a) A
    symbolic representation of the classical Post machine (a
    single tape).\\
    (b) A symbolic representation of computations with a quantum
    superposition of tapes. The programme contains new types of
    instructions but it is still classical.}
  \label{fig:post}
\end{figure}
A state of the machine is described by two independent components: the
tape (data) and the instruction list (programme), see
Fig.~\ref{fig:post}(a).  The tape is assumed to be an infinite
sequence of cells with only a finite number of them holding mark
\(1\), all others assumed to be ``empty'' (holding \(0\)). Another
important property of the tape 
is the \emph{current
  cell} for observation/modification pointed by a reading head.  

The second machine's component---\emph{programme}---is a finite list
of instructions with the second pointer marking the current command. The
statements are taken from a very limited set and request modifications
of the current tape's cell or respective movements of the reading head
and the instruction pointer.

\begin{rem}
  \label{re:duality}
  The division 
  into ``data tape'' and ``programme'' seems to be a fundamental one.
  This duality is reflected in both---architectures of modern
  computers and the computer science paradigm of ``Algorithms and Data
  Structures''~\cite{WirthAlgorithsDS}.
\end{rem}

A typical quantum computation~\citelist{\cite{Shor94} \cite{Grover96}}
can be modelled by a \emph{quantisation} of the tape in a machine, see
Fig.~\ref{fig:post}(b).  This means that instead of a classical tape
holding a sequence of classical bits one considers a quantum tape: a
finite number of cells holding \emph{qubits}. Qubits are assumed to be
able to store linear combinations (superpositions) of values \(0\) and
\(1\).

A quantisation of the other half---the programme---is rarely
considered: it is still a linear sequence of corresponding
instructions, which are unitary operators on qubits in this case. Thus
a common quantum computer is strictly speaking semi-classical or
quantum-classical computer only. To get fully quantised computer one
can additionally request superpositions of computer states and/or
programmes. However a realisation of superposition for instructions
can be confusing.

In this paper we consider an alternative approach. Firstly we get
unification of the tape and the reading head position into a single
coordinate. Computer's programme is linked to the another coordinate.
Then we can quantise it in a single move. Computational speed of such
a computer cannot be directly compared to a classical one, since it
will not only process data in parallel but also perform different
computational stages at the same time.

\section{Phase Space Computations and Hamiltonian Dynamics}
\label{sec:phase-space-comp}

To obtain the dynamical description of computations we blend
the state of the tape and position of the reading head into a
single parameter. We interpret the finite sequence of \(1\)'s and
enclosed among them \(0\)'s on the tape as a
\wiki{Dyadic_fraction}{dyadic rational} number with the binary point
at the immediate right to the current cell. Then the standard
actions of the reading head (the first column of Tab.~\ref{tab:head})
can be translated into operations on the 
set \(\Space{D}{}\) of dyadic numbers (the second column of Tab.~\ref{tab:head}).
\begin{table}[htbp]
  \centering
  \begin{tabular}{||c|c|c||}
    \hline\hline
    Head action & Arithmetic operation\strut & Value of
    \(\Delta_p H(q_0,p_0)\)\\
    \hline\hline
    Head to the left & Divide the fraction by \(2\)&
    \(-\frac{1}{2}q_0\)\strut \\
    \hline
    Head to the right & Multiply the fraction by \(2\) & \(q_0\)\\
    \hline
    Replace \(0\) by \(1\) 
    & Add \(1\) to the fraction & \(1\)\\
    \hline
    Replace \(1\) by \(0\) 
    & Subtract  \(1\) from the fraction & \(-1\)\\
    \hline\hline
  \end{tabular}
  \caption[Reading head actions, arithmetic and Hamiltonian]{The first
    column lists actions of the reading head of a machine, the second
    column translates them into dyadic arithmetic. The third column
    provides values of a Hamiltonian which direct those
    transformations.} 
  \label{tab:head}
\end{table}

Similarly the set \(\Space[n]{Z}{}=\{1, 2, \ldots, n\}\) can index a
programme of \(n\) instructions. Thus a full state of a machine is
described by a point \((q,p)\in\Space{D}{}\times\Space[n]{Z}{}\).
Calculation is a dynamic on this set with a discrete time parameter
\(t\in\Space{N}{}\). An iteration from a current state \((q_t,p_t)\)
to the next one \((q_{t+1},p_{t+1})\) is given by the pair of finite
differences equations:
\begin{equation}
  \label{eq:hamilton-calc}
  \Delta_t q = \Delta_p H, \qquad   \Delta_t p = -\Delta_q H.
\end{equation}
Here \(\Delta_q H\) and \(\Delta_p H\) is a pair of functions
\(\Space{D}{}\times\Space[n]{Z}{}\rightarrow \Space{Z}{}\) and
\(\Space{D}{}\times\Space[n]{Z}{}\rightarrow \Space{D}{}\)
respectively. The function \(\Delta_p H\) defines transformations of
the tape according to the third column in Tab.~\ref{tab:head}. The
programme flow is directed by \(\Delta_q H\) as described in
Tab.~\ref{tab:instruction}. 
\begin{rem}
  We intentionally use notations resembling Hamiltonian dynamics in
  order to exploit the duality between data and algorithms mentioned
  in Rem.~\ref{re:duality}. However the exact mathematical formalism
  for this duality is still missing. For example, canonical
  transformations mixing data and programme can be related to the
  philosophy behind \wiki{Prolog}{\texttt{Prolog}} and
  \wiki{Lisp_\%28programming_language\%29}{\texttt{Lisp}} programming languages.
\end{rem}

\begin{table}[htbp]
  \centering
  \begin{tabular}{||c|c||}
    \hline\hline
    Instruction pointer & Value of \(\Delta_q H(q_0,p_0)\)\\
    \hline\hline
    Next Instruction & \(-1\)\\
    \hline    
    Go to \(p_1\) & \(p_0-p_1\)\\
    \hline
    If cell is \(1\) go to \(p_1\) & \(
    \left\{ \begin{array}{ll}
        p_0-p_1,& \text{ if } [q_0]=1\mod 2;\\
        1, & \text{ if } [q_0]\neq 1\mod 2.
      \end{array}\right.
    \)  
    \\
    \hline\hline
  \end{tabular}
  \caption[Instructions and Hamiltonian]{Movements of the programme
    pointer (the first column) and the corresponding values of a
    Hamiltonian (the second column).}
  \label{tab:instruction}
\end{table}

\begin{thm}
  Calculations of a Post machine is described by a discrete dynamics in
  the phase space \(\Space{D}{}\times \Space[n]{Z}{}\) defined by the
  equations~\eqref{eq:hamilton-calc}. A 
  programme corresponds to a Hamiltonian governing the dynamic.
\end{thm}
\begin{table}[htbp]
  \centering
    \begin{tabular}{||c|c||}
      \hline\hline
      Abstract computing & Hamilton dynamic\\
      \hline\hline
      Tape state & Coordinate\\
      \hline
      Inner state & Momentum\\
      \hline
      Program & Hamiltonian\\
      \hline
      Execution & Dynamics\\
      \hline
      Inclusion-Exclusion & Wave superposition\\
      \hline\hline
    \end{tabular}
  \caption[Calculation and Dynamics]{The correspondence between
    element of abstract calculations and dynamics in the phase space.}
  \label{tab:dynamics}
\end{table}
Tab.~\ref{tab:dynamics} shows a correspondence between notions of
computing and Hamilton dynamics.  Developing this approach we can
define a \emph{fully} quantum computation through quantisation of the
classical discrete dynamics.  This gives simultaneous propagation
along all possible paths, which means parallel procession of data
\emph{and} the programme similarly.

\section{Example: Polynomial Sequences of Binomial Type}
Classical computations of many combinatorial quantities is based on
the \wiki{Inclusion-exclusion_principle}{in\-clus\-ion-ex\-clus\-ion
  principle}~\cite{StanleyI}*{\S~2.1}. Its quantum counterpart is the
superposition of wave functions: the resulting probability can be
anything from the sum (inclusion) to the difference (exclusion) of
given probabilities. Thus such combinatorial calculations are very
suitable for quantum computations.

For example, let \(q_n(x)\) be a \emph{token}~\citelist{\cite{Kisil01b}
  \cite{Kisil97b}} from \(\Space{N}{}\) to \(\Space{R}{}\), i.e. the
sequence of polynomials of \(\deg q_n = n\) satisfying to the identity:
\begin{displaymath}
  q_n(x+y)=\sum_{k=0}^n q_{k}(y) q_{n-k}(x),
\end{displaymath}
If \(q_n(x)\) is such a token then a polynomial sequence \(p_n(x)=n!
q_n(x)\) is of \emph{binomial type}~\cite{RotaWay}*{\S~4.3}. Examples
are provided by power monomials, falling (rising) factorials, Abel,
Laguerre and many other famous polynomials.

A dynamics in a configurational space \(Q\) can be described by the
\emph{propagator} \(K(q_2,t_2;q_1,t_1)\)---a complex valued function
defined on \(\Space{Q}{} \times \Space{R}{} \times \Space{Q}{} \times
\Space{R}{}\).  It is a probability amplitude for a transition \(q_1
\rightarrow q_2\) from a state \(q_1\) at time \(t_1\) to \(q_2\) at
time \(t_2\).  
The fundamental assumption about the quantum world is the
\emph{absence of trajectories} for a system's evolution through the
configurational space \(\Space{Q}{}\): the system at any time \(t_i\)
could be found at any point \(q_i\).  

\person{R.~Feynman} developing ideas of \person{A.~Einstein},
\person{M.V.~Smoluhovski} and \person{P.A.M.~Dirac} proposed an expression
for the propagator via the ``integral over all possible paths'':
\begin{displaymath}
  K(q_2,t_2;q_1,t_1)=\int \frac{\mathcal{D}q\, \mathcal{D}p}{h} \exp
    \left( 
    \frac{i}{\hbar} \int\limits_{t_1}^{t_2} dt \left(p\dot{q}-H(p,q)
    \right) \right)
  .
\end{displaymath} Here \(H(p,q)\) is the 
Hamiltonian of the system
.  The inner integral is 
over a path in the phase space. The
outer integral is taken over ``all possible paths between two given
points with respect to a measure \(\mathcal{D}q\, \mathcal{D}p\) on
paths in the phase space''.
\begin{prop}
  Any quantum system is a quantum computer for an evaluation of its
  own propagator \(K\), computation is done simultaneously along all
  possible paths.
\end{prop}

In this way we obtain the path computation
formula for polynomials \(q_n\)~\cite{Kisil98d}:
\begin{displaymath}
  q_n(x)= \int\!\mathcal{D}k\mathcal{D}p\,
  \exp\!\!\int\limits_0^x
  (-ipk'+h(p))\,dt, \qquad \text{ where } h(p)=\sum_{k=0}^\infty q_k'(0)
  e^{ipk}.
\end{displaymath} 
Thus a quantum system with the above Hamiltonian \(h(p)\) allows to calculates
\(q_n\) in a single operation (measurement). This looks
unrealistically quick and one can ask: how to
compare speeds of quantum and classical computations after all?

\section{Quantum Computers with Classical Terminals}
\label{sec:quant-comp-with}

A discussion of quantum computers is often limited to quantum
algorithms.  However this an oversimplification, which does not
include the process of qubit preparation (input of data), building
sequences of quantum gates (programming) and reading of the final
state (data output). 
Of course, in the
classical case these three processes can be done in a negligible time
in comparison with the actual computation. However, this is no longer
true for quantum computations.
\begin{example} Let us review two most known quantum algorithms.
  \begin{enumerate}
  \item Shor's factorisation algorithm~\cites{Shor94} required the
    quantum circuit to be reassembled accordingly every time a new
    random number was chosen for a test. Thus the time of circuit
    assembling (programming) should be included in the overall
    computational cost.
  \item Grover's database search algorithm~\cites{Grover96} requires
    several repeated recalculations, each of which would destroy the
    database (the projection postulate of quantum
    measurement~\citelist{\cite{Mackey63} 
    }). Thus the
    time for rebuilding a database (data input) and measurement (data
    output) should be included in the overall computational cost.
  \end{enumerate}
\end{example}

For more realistic consideration we have to add \emph{classical
interfaces} for input and output to make quantum computations really usable.
At present even a simple quantum step like two qbits swapping 
is done by a millions of classical computational steps.
Is it a \emph{present day} technological limitation or \emph{fundamental} exchange
rate between cost of a quantum and classical computation?
If \emph{an application} of an existent  quantum gate is so expensive, how expensive is
\emph{to built} a case-specific quantum circuit for \(f(x)=a^x\)~\cite{Grover96} or
quantum Fourier transform~\cite{Shor94}? Such questions are already hinted
in~\cite{Shor94} but are rarely discussed in depth. Consequently we
miss not only clear answers but even the understanding of their importance.

In the first half of this paper we presented classical and quantum
computations as dynamics.  Then a quantum computer with classical
terminals shall be represented by a dynamics of a quantum-classical
aggregate system.
Is there  a consistent theory to describe such a dynamics? 
This is a debated topics with the majority of physicists believing
that this is fundamentally impossible~\citelist{\cite{CaroSalcedo99}
  \cite{Sahoo04}}. If this is so, shall it be interpreted as our inability (as a
macroscopic and thus classical objects) to efficiently interact with quantum
computing devices even if they are to be built?

Quantum-classical dynamics is oftenly connected with an existence
of special quantum-classic bracket which shall unify (and replace)
both quantum commutator and Poisson brackets.
A mathematical model for a classical system attached as an
input/output terminal to a quantum computer can be attempted from the
quantum-classical formalism proposed in~\citelist{\cite{Kisil02e}
  \cite{BrodlieKisil03a} \cite{Kisil05c} \cite{Kisil09a}}.  Such a
model would provide an opportunity for effective estimation of the
overall cost of quantum computing during the entire cycle:
preparation-computation-reading. 

To stimulate an attention to this issue we wish to conclude
by the following:
\begin{conj}[``Golden rule'' of quantum-classic information]
  A gain in quantum algorithms is outweighed by losses in classical I/O and programing. 
\end{conj}

\small
\bibliography{abbrevmr,akisil,analyse,algebra,arare,aclifford,aphysics,acompute,acombin}

\newcommand{\noopsort}[1]{} \newcommand{\printfirst}[2]{#1}
  \newcommand{\singleletter}[1]{#1} \newcommand{\switchargs}[2]{#2#1}
  \newcommand{\irm}{\textup{I}} \newcommand{\iirm}{\textup{II}}
  \newcommand{\vrm}{\textup{V}} \providecommand{\cprime}{'}
  \providecommand{\eprint}[2]{\texttt{#2}}
  \providecommand{\myeprint}[2]{\texttt{#2}}
  \providecommand{\arXiv}[1]{\myeprint{http://arXiv.org/abs/#1}{arXiv:#1}}
  \providecommand{\CPP}{\texttt{C++}} \providecommand{\NoWEB}{\texttt{noweb}}
  \providecommand{\MetaPost}{\texttt{Meta}\-\texttt{Post}}
  \providecommand{\GiNaC}{\textsf{GiNaC}}
  \providecommand{\pyGiNaC}{\textsf{pyGiNaC}}
  \providecommand{\Asymptote}{\texttt{Asymptote}}
\begin{bibdiv}
\begin{biblist}

\bib{BrodlieKisil03a}{incollection}{
      author={Brodlie, Alastair},
      author={Kisil, Vladimir~V.},
       title={Observables and states in {$p$}-mechanics},
        date={2003},
   booktitle={Advances in mathematics research. vol. 5},
   publisher={Nova Sci. Publ.},
     address={Hauppauge, NY},
       pages={101\ndash 136},
        note={\arXiv{quant-ph/0304023}},
      review={\MR{MR2117375}},
}

\bib{CaroSalcedo99}{article}{
      author={Caro, J.},
      author={Salcedo, L.~L.},
       title={Impediments to mixing classical and quantum dynamics},
        date={1999},
     journal={Phys. Rev.},
      volume={A60},
       pages={842\ndash 852},
        note={\arXiv{quant-ph/9812046}},
}

\bib{Grover96}{inproceedings}{
      author={Grover, Lov~K.},
       title={A fast quantum mechanical algorithm for database search},
        date={1996},
   booktitle={Proceedings of the twenty-eighth annual acm symposium on the
  theory of computing ({Philadelphia, PA}, 1996)},
   publisher={ACM},
     address={New York},
       pages={212\ndash 219},
      review={\MR{MR1427516}},
}

\bib{Kisil97b}{article}{
      author={Kisil, Vladimir~V.},
       title={Umbral calculus and cancellative semigroup algebras},
        date={2000},
        ISSN={0232-2064},
     journal={Z. Anal. Anwendungen},
      volume={19},
      number={2},
       pages={315\ndash 338},
        note={\arXiv{funct-an/9704001}. \Zbl{0959.43004}},
      review={\MR{MR1768995 (2001g:05017)}},
}

\bib{Kisil98d}{article}{
      author={Kisil, Vladimir~V.},
       title={Polynomial sequences of binomial type and path integrals},
        date={2002},
        ISSN={0218-0006},
     journal={Ann. Comb.},
      volume={6},
      number={1},
       pages={45\ndash 56},
        note={\arXiv{math/9808040} \Zbl{1009.05013}},
      review={\MR{2003e:05010}},
}

\bib{Kisil01b}{incollection}{
      author={Kisil, Vladimir~V.},
       title={Tokens: an algebraic construction common in combinatorics,
  analysis, and physics},
        date={2002},
   booktitle={Ukrainian mathematics congress---2001 (ukrainian)},
   publisher={Nats\=\i onal. Akad. Nauk Ukra\"\i ni \=Inst. Mat.},
     address={Kiev},
       pages={146\ndash 155},
        note={\arXiv{math.FA/0201012}},
      review={\MR{MR2228860 (2007d:05010)}},
}

\bib{Kisil02e}{article}{
      author={Kisil, Vladimir~V.},
       title={{$p$}-mechanics as a physical theory: an introduction},
        date={2004},
        ISSN={0305-4470},
     journal={J. Phys. A},
      volume={37},
      number={1},
       pages={183\ndash 204},
        note={\arXiv{quant-ph/0212101},
  \href{http://stacks.iop.org/0305-4470/37/183}{On-line}. \Zbl{1045.81032}},
      review={\MR{MR2044764 (2005c:81078)}},
}

\bib{Kisil05c}{article}{
      author={Kisil, Vladimir~V.},
       title={A quantum-classical bracket from {$p$}-mechanics},
        date={2005},
        ISSN={0295-5075},
     journal={Europhys. Lett.},
      volume={72},
      number={6},
       pages={873\ndash 879},
        note={\arXiv{quant-ph/0506122},
  \href{http://dx.doi.org/10.1209/epl/i2005-10324-7}{On-line}},
      review={\MR{MR2213328 (2006k:81134)}},
}

\bib{Kisil09a}{article}{
      author={Kisil, Vladimir~V.},
       title={Comment on `do we have a consistent non-adiabatic
  quantum-classical mechanics?'},
        date={submitted},
     journal={Europhys. Lett. EPL},
        note={\arXiv{0907.0855}},
}

\bib{KnuthACP250}{book}{
      author={Knuth, Donald~E.},
       title={The art of computer programming},
     edition={Second},
   publisher={Addison-Wesley Publishing Co., Reading, Mass.-London-Amsterdam},
        date={1975},
        note={Volume 1: Fundamental algorithms, Addison-Wesley Series in
  Computer Science and Information Processing},
      review={\MR{MR0378456 (51 \#14624)}},
}

\bib{RotaWay}{book}{
      author={Kung, Joseph P.~S.},
      author={Rota, Gian-Carlo},
      author={Yan, Catherine~H.},
       title={Combinatorics: the {R}ota way},
      series={Cambridge Mathematical Library},
   publisher={Cambridge University Press},
     address={Cambridge},
        date={2009},
        ISBN={978-0-521-73794-4},
      review={\MR{MR2483561}},
}

\bib{Mackey63}{book}{
      author={Mackey, George~W.},
       title={Mathematical foundations of quantum mechanics},
   publisher={W.A.~Benjamin, Inc.},
     address={New York, Amsterdam},
        date={{\noopsort{}}1963},
}

\bib{Sahoo04}{article}{
      author={Sahoo, Debendranath},
       title={Mixing quantum and classical mechanics and uniqueness of
  {Planck's} constant},
        date={2004},
     journal={J. Phys. A: Math. Gen.},
      volume={37},
       pages={997\ndash 1010},
        note={\arXiv{quant-ph/0301044}},
}

\bib{Shor94}{incollection}{
      author={Shor, Peter~W.},
       title={Algorithms for quantum computation: discrete logarithms and
  factoring},
        date={1994},
   booktitle={35th annual symposium on foundations of computer science (santa
  fe, nm, 1994)},
   publisher={IEEE Comput. Soc. Press},
     address={Los Alamitos, CA},
       pages={124\ndash 134},
      review={\MR{MR1489242}},
}

\bib{StanleyI}{book}{
      author={Stanley, Richard~P.},
       title={Enumerative combinatorics. {V}ol. 1},
   publisher={Cambridge University Press},
     address={Cambridge},
        date={1997},
        ISBN={0-521-55309-1},
        note={With a foreword by Gian-Carlo Rota, Corrected reprint of the 1986
  original},
      review={\MR{98a:05001}},
}

\bib{WirthAlgorithsDS}{book}{
      author={Wirth, Niklaus},
       title={Algorithms and data structures},
     edition={Revised},
   publisher={Prentice Hall International},
     address={Englewood Cliffs, NJ},
        date={1986},
        ISBN={0-13-022005-1},
      review={\MR{MR808586 (87b:68001)}},
}

\end{biblist}
\end{bibdiv}
\end{document}